\documentstyle[preprint,eqsecnum,aps]{revtex}
\begin{document}
\draft
\input{epsf.tex}

\def\Rl{{\mathchoice
{\setbox0=\hbox{$\displaystyle\rm R$}\hbox{\hbox to0pt
{\kern0.4\wd0\vrule height0.9\ht0\hss}\box0}}
{\setbox0=\hbox{$\textstyle\rm R$}\hbox{\hbox to0pt
{\kern0.4\wd0\vrule height0.9\ht0\hss}\box0}}
{\setbox0=\hbox{$\scriptstyle\rm R$}\hbox{\hbox to0pt
{\kern0.4\wd0\vrule height0.9\ht0\hss}\box0}}
{\setbox0=\hbox{$\scriptscriptstyle\rm R$}\hbox{\hbox to0pt
{\kern0.4\wd0\vrule height0.9\ht0\hss}\box0}}}}
\def\L{{\rm L}}

\tighten

\preprint{\vbox{
\rightline{SU-GP-99/3-2}
\rightline{NSF-ITP-99-14}
\rightline{hep-th/9903213}}}
\title{Brane Baldness vs. Superselection Sectors}
\author{Donald Marolf}
\address{Institute for Theoretical Physics, University of California,
Santa Barbara, CA 93106}
\address{Physics Department, Syracuse University, Syracuse,
         NY 13244}
\author{Amanda Peet}
\address{Institute for Theoretical Physics, University of California,
Santa Barbara, CA 93106}

\maketitle

\begin{abstract}
The search for intersecting brane solutions in supergravity is a large
and profitable industry.  Recently, attention has focused on finding
localized forms of known `delocalized' solutions.  However, in some
cases, a localized version of the delocalized solution simply does not
exist.  Instead, localized separated branes necessarily delocalize as
the separation is removed.  This phenomenon is related to black hole
no-hair theorems, i.e. `baldness.'  We continue the discussion of this
effect and describe how it can be understood, in the case of Dirichlet
branes, in terms of the corresponding intersection field theory.  When
it occurs, it is associated with the quantum mixing of phases and lack
of superselection sectors in low dimensional field theories.  We find
surprisingly wide agreement between the field theory and supergravity
both with respect to which examples delocalize and with respect to the
rate at which this occurs.
\end{abstract}

\vfil

\begin{center}{March, 1999}\end{center}

\eject

\section{Introduction}
\label{intro}

It is a well-known fact that, while string perturbation theory in
intersecting D-brane backgrounds is discussed in terms of localized
branes\cite{JP}, classical supergravity solutions representing such
localized branes are typically difficult to construct.  One often
works with supergravity solutions where the branes are less well
localized than one might like, and in fact have extra translational
symmetries \cite{l1} - \cite{l14.2}; see \cite{YR} for a thorough
review. One says that the branes have been `smeared' over some of the
directions that were, a priori, transverse to their world volume.  In
some cases, with a bit more work, one can construct explicit fully
localized solutions, at least in the `near-core' limit \cite{ITY,aki}.

However, there are some situations where fully localized supergravity
solutions of the desired type simply do not exist.  This happens when
the world volume of one type of brane (B) is contained inside that of
another (A), or inside the volume through which brane A has been
smeared, and when the branes intersect on a manifold of sufficiently
low dimension; the details will be explained below.  If the branes are
separated in a direction transverse to both types (A and B), then
localized solutions exist.  Nonetheless, as the transverse separation
between the branes is removed, the type B brane delocalizes.  This
phenomenon does not occur when the dimension of the intersection
manifold is sufficiently high.

This delocalization was studied in \cite{SM} for one-branes parallel
to five-branes.  As discussed there, this effect is related to a black
hole no-hair theorem.  Black hole no-hair results tell us that, in
certain cases, black hole horizons must be uniform.  A pertinent
example occurs in Einstein-Maxwell theory: when an electric charge $q$
is brought near a black hole, the charge $q$ appears to be delocalized
over the black hole horizon \cite{membrane}.  The situation discussed
below is similar as, when delocalization occurs, the charge of the
type B brane appears to be delocalized over the entire horizon of the
type A brane.  For such cases, the type A branes are `bald' and unable
to support such hair.

Now, recall that there are certain dualities between supergravity and
brane gauge theories, as in the AdS/CFT correspondence \cite{Juan} and
its $p$-brane generalizations \cite{IMSY}; see also e.g. \cite{BooPS}.  
The general picture of a
duality is that there is a single quantum theory which has several
distinct classical limits.  In our case, one of these limits would
give classical supergravity while the other would give classical brane
gauge theory.  This means that aspects of the dynamics which are
classical from the point of view of one theory correspond to strongly
quantum mechanical effects in terms of the other.  We will further
generalize the AdS/CFT limit to describe interacting A and B branes,
and we will see that the delocalization of classical supergravity
corresponds to large fluctuations of a modulus field on the gauge
theory side.  In particular, its occurrence is related to the quantum
mixing between phases and the lack of superselection sectors
associated with the asymptotic values of massless fields in low
dimensional quantum field theories; i.e., on the gauge theory side,
delocalization is controlled by the Coleman-Mermin-Wagner
theorem\cite{MW,SC}.  The relevant quantum mixing is between the gauge
theory phase which describes type A branes separated from type B
branes in an overall transverse direction, which we will refer to
generally as the `separated' branch, and the phase that describes such
branes with no transverse separation, which we call the `coincident'
branch\footnote{This result is not, in fact, in conflict with
\cite{Wit,ABKSM}, as we will discuss.}  That some correspondence of
this sort should occur was suggested in \cite{Pelc}.  The consistency
of this picture will be explained in section
\ref{IFT}.  We will see that, not only will this interpretation
successfully predict the cases for which the type B branes delocalize,
it will also give the correct rate at which the delocalization occurs
as the transverse separation is removed.

Before progressing to the main paper, it is perhaps worthwhile to
display an explicit and tractable example of the supergravity
delocalization phenomenon.  It turns out that this is much easier to
see for a D-instanton located near a D3-brane than for the systems
studied in \cite{SM}.  The point is that the classical supergravity
solution for a localized BPS D-instanton in AdS${}_5 \times$S${}^5$ is
known in terms of elementary functions \cite{CHW,BGKR,srd} while, for
the other examples, the supergravity solution is known only as a
(convergent) infinite series.

The near D3-brane geometry is just AdS${}_5\times$S${}^5$.  For
simplicity, we take the case \cite{CHW} where the D-instanton is
smeared over the S${}^5$, but the solution localized on the S${}^5$ is
also known explicitly \cite{BGKR}.  By writing the
AdS${}_5\times$S${}^5$ metric in Poincar\'e coordinates,
\begin{equation}
\label{metric}
ds^2 = R^2 (U^{-2} dU^2 + U^2 \delta_{ij} dy^i dy^j + d\Omega_5^2),
\end{equation}
we may interpret the solution of \cite{CHW} as the near-core part of
an asymptotically flat (Euclidean) spacetime with a spherical
(S${}^5$) shell of D-instantons located near the three-branes.  In
(\ref{metric}), $R$ is the radius of the AdS${}_5$ and of the S${}^5$.
In the presence of the D-instantons, the (Einstein) metric is
unchanged, and the dilaton is given by
\begin{equation}
e^\phi = c_0 + c_1 {{(d^2 +2) (d^4 + 4d^2 -2)} \over {d^3 (d^2
+4)^{3/2}}},
\end{equation}
where
\begin{equation}
\label{d}
d^2 = U_0 U \left[ (U^{-1} - U_0^{-1})^2 + \sum_{i=0}^3 (y^i -
y^i_0)^2
\right],
\end{equation}
with $(U_0,y^i_0)$ the location of the D-instantons.  This setup
allows a separation between the D-instantons and the origin of
AdS${}_5$.  Note that far from the D-instantons $e^\phi$ tends to a
finite and nonzero constant, $c_0+c_1$.  The constant $c_0$ is
arbitrary while the constant $c_1$ is proportional to the instanton
charge.  Consider now the limit in which the instantons are moved onto
the three-brane; that is, the limit in which $U_0 \rightarrow 0$ with
fixed $c_0,c_1, y^i_0$. Note that $U,y^i$ should also remain fixed as
we wish to examine the solution at a given location relative to the
three-brane.  In this limit, (\ref{d}) diverges so that we have
\begin{equation}
\left( e^\phi\right)_{\rm lim} = c_0 + c_1.
\end{equation}
Therefore the dilaton is now constant (and it is in fact the same
constant as the asymptotic value of $e^\phi$ at generic D-instanton
position).  In particular, the solution no longer has any dependence
on the coordinates $y^i$ along the three-brane.  We see that the
D-instantons have delocalized as they were moved toward the
three-brane.  Note that this does not happen suddenly at $U_0 =0$, but
rather gradually.  As viewed from a fixed point relative to the
three-brane, the field created by the instantons smoothly blurs out as
we decrease $U_0$ to zero.  In this context, we see that our
delocalization is related to the well-known scale/radius duality of
this system \cite{CHW,BGKR,srd}.  
The behavior of the D-instanton solution \cite{BGKR} that is
localized on the S${}^5$ as well as in the AdS${}_5$ is similar.
Much the same behavior was seen in \cite{SM} for a localized one-brane
as it is moved onto a localized five-brane and, as we will see below,
it occurs in many other cases as well.

The outline of this paper is as follows.  In section \ref{SUGRAV}, we
extend the analysis of \cite{SM} to consider the full range of two
charge intersecting brane solutions whose mathematical form is similar
to that of the localized D1-branes parallel to D5-branes
(D1$\parallel$D5).  For example, the analysis includes the localized
D0$\parallel$D4 system, and the D2$\parallel$D6 system studied in
\cite{ITY}.  We will see that the D0$\parallel$D4 system delocalizes
as does the D1$\parallel$D5 system, but that the same methods find the
D2$\parallel$D6 system to remain localized.  This is a useful check of
the method and resolves the apparent conflict\footnote{In
\cite{SM}, it was stated that there was a normalization problem with 
the solutions of \cite{ITY}.  This is not, in fact, correct. For
completeness, the normalizations are discussed in appendix \ref{Y}.}
between the results of \cite{SM} for the D1$\parallel$D5 system and
the explicit construction of the localized D2$\parallel$D6 solutions
in \cite{ITY}.  The D2$\parallel$D6 system is simply different from
the D1$\parallel$D5 system, in a way that will be discussed below.
Other solutions considered below are similar to a D1-brane
orthogonally intersecting a D3-brane, where (say) the D1-brane is
smeared along the D3-brane world volume.  In this case, we will see
that when the transverse separation between a (localized) D3 brane and
the smeared D1 branes is removed, the D3 brane delocalizes along the
D1-brane world volume.  To be quite general, we will allow arbitrary
transverse separation between the two types of branes, although we
will still refer to these solutions as `intersecting' brane
spacetimes.  For many of these systems, certain `near-core' solutions
were constructed in \cite{Youm}; we argue in appendix \ref{Y} that,
while these are certainly valid solutions to the supergravity
equations, due to the subtleties of boundary conditions they are not
the appropriate ones to consider in our context.

After studying the supergravity solutions, we turn in section
\ref{IFT} to a discussion of the corresponding intersection field
theories and the appropriate AdS/CFT limit \cite{IMSY}.  We will see
that our delocalization phenomenon (when it occurs!)  corresponds to a
quantum mixing between phases of the Yang-Mills theory and to the fact
that 0+1 and 1+1 field theories are not superselected by the
asymptotic values of the massless fields.  Finally, we discuss a few
remaining issues in section \ref{disc}.

\section{Delocalization in Supergravity}
\label{SUGRAV}

Here, we consider BPS solutions with two types of branes, A and B,
each localized at fixed, but different, values of isotropic
coordinates $x_\perp$ which label the space transverse to both
branes\footnote{ We describe locations in the spacetime in terms of
isotropic coordinates.  In all cases below, this may be translated
into a coordinate invariant statement by referring, for example, to
the surface in spacetime on which the various gauge field strengths
take a given value.  However, it is simpler to discuss the solutions
directly in terms of isotropic coordinates.}.  That is, we consider
BPS solutions with the branes separated in the transverse direction.
The branes may carry Ramond-Ramond or Neveu-Schwarz--Neveu-Schwarz
charge in Type IIA/B string theory or be M-branes, and the A- and
B-branes need not be of the same sort.  For example, the type A brane
may be R-R while the type B branes are NS-NS.  Thus, we include the
case of a fundamental string intersecting (but not ending on) a
D-brane.  The equations of motion for all of these cases have a
similar structure as they are related by T- and S- duality.  A similar
analysis also applies to Kaluza-Klein monopoles and various sorts of
waves.  In this section, we refer to all such cases as `branes.'

The world-volumes of the type A and B branes will in general have
certain directions in common.  We label the directions common to both
by coordinates $t,z_I$.  Here, the $I$ is a label denoting the
`intersection' directions.  Similarly, we denote by $z_a$ the
worldvolume directions of brane A not in I, and we denote by $z_b$ the
worldvolume directions of brane B not in I.  We take brane A to be
smeared in the $z_b$ directions, as well as in any directions in which
brane B has been smeared.  We will have no need to refer explicitly to
directions in which both branes have been smeared, as these can be
removed by a T-duality symmetry transformation, but we label
directions in which brane A (but not B)\footnote{We could
alternatively smear brane B, but this will not affect the
delocalization behavior.} has been smeared by $w$.  Finally, $x_\perp$
labels directions orthogonal to both branes along which neither brane
has been smeared.  Our conventions are conveniently summarized by the
following table, where ($\bullet$) denotes a direction along a brane,
($\equiv$) denotes a direction in which a brane has been smeared, and
an empty space denotes a direction orthogonal to the brane in which it
has not been smeared.
\begin{center}
\begin{tabular}{|c||c|c|c|c|c|c|} \hline
{}   \ & \  $t$       \ & \  $z_I$     \ & \  $z_a$     \ & \  $z_b$
  \ & \  $w$
  \ & \  $x_\perp$
        \\ \hline
$A$ \ & \  $\bullet$ \ & \  $\bullet$ \ & \  $\bullet$ \ & \
$\equiv$  \
& \  $\equiv$   \
& \  {}
        \\ \hline
$B$ \ & \  $\bullet$ \ & \  $\bullet$ \ & \  {}        \ & \
$\bullet$ \
& \   {}  \
& \  {}
        \\ \hline
\end{tabular}
\end{center}
We assume that the $z_a$, $z_b$, and $w$ directions are compactified
to form tori of volumes $V_a$, $V_b$, and $V_w$ respectively (the
noncompact case just corresponds to the $V\rightarrow\infty$ limit).
The $z_I$ directions are taken to be uncompactified, or compactified
on a manifold of very large volume.  We will also use the symbols
$a,b,w$ to denote the number of $z_a, z_b, w$ coordinates.  Our
discussion also applies to BPS branes at angles, though we will not
consider this case explicitly.

Familiar examples of this class are D$(p-4)$-branes (B) oriented
parallel to D$p$-branes (A) as well as various intersecting brane
solutions \cite{l8,l9,l14,YR} with one brane (A) smeared along the
world-volume directions of the other brane (B).  Additional cases are
generated by further smearing.  Thus, we may refer to brane A as `the
bigger brane,' in the sense that the {\it smeared} type A branes fill
out a higher dimensional volume.

For BPS configurations we make the ansatz
\begin{equation}\label{intbraneg}
ds^2 = {{1}\over{\sqrt{H_AH_B}}}(-dt^2 + dz_I^2) +
\sqrt{{H_A}\over{H_B}}(dz_b)^2 + \sqrt{{H_B}\over{H_A}}(dz_a)^2 +
\sqrt{H_AH_B}(dx_\perp^2) \,,
\end{equation}
and
\begin{equation}
e^\Phi = g_s H_A^{(3-A)/4} H_B^{(3-B)/4} \,.
\end{equation}
We have not written the gauge fields explicitly, but they have the
standard form.  Constructing the supergravity solutions then reduces
(see, e.g., \cite{l9,l10,l12,HM3}) to solving the following equations
of motion:
\begin{equation} \label{A}
\partial_{x_\perp}^2 H_A (x_\perp) = {{q_A} \over {V_b V_w}}
\delta(x_\perp),
\end{equation}
\begin{equation}
\label{B}
(\partial_{x_\perp}^2 + \partial_w^2) H_B(x_\perp,w,z_{a})
+  H_A (x_\perp)  \partial_{z_{a}}^2  H_B(x_\perp,w,z_{a})  =  q_B
\delta (x_\perp, x_{\perp 0}) \delta(w,w_0)
\delta (z_{a}, z_{a 0}),
\end{equation}
where $\partial_{x_\perp}^2$ represents the flat space Laplacian in
the $x_\perp$ directions, and similarly for $\partial_w^2$ and
$\partial_{z_{a} }^2$.  Note that, by construction, the solution has
no dependence on $t$, $z_I$ or $z_b$.  The delta-function sources,
each of the appropriate dimension, ensure that the solution carries
gauge field fluxes corresponding to charges $q_A$, $q_B$ at the
locations specified.  Note that we have taken brane A to lie at the
origin of the transverse coordinates $x_\perp$, while we have placed
brane B at the location $(x_{\perp 0}, w_0, z_{a 0})$.  We have chosen
to let $q_A,q_B$ denote the total charge of the type A and B branes,
although only the charge density of the type A branes appears in the
above equations of motion (\ref{A}).

The first equation of motion (\ref{A}) is just Laplace's equation.
Solutions appropriate to asymptotically flat boundary conditions may
be found if the $x_\perp$ coordinates label a $d$ dimensional space
with $d \ge 3$, and we confine ourselves to this case.  The type A
branes are associated with a `charge radius' $r_A$ proportional to
$q_A^{1/(d+b+w-2)}$, but the behavior of the supergravity solution is
controlled by the length scale
\begin{equation}
\hat r_A = ({{r_A^{d+b+w-2}} \over {V_bV_w}})^{1/(d-2)} \,.
\end{equation}
The appropriate solution may thus be written
\begin{equation}
\label{HA}
H_A (x_\perp) = 1 + {{\hat r_A^{d -2}} \over {r_\perp^{d-2}}},
\end{equation}
where $r_\perp = |x_\perp|$.

The method of \cite{SM} then uses the symmetries and the linearity of
equation (\ref{B}) to solve for $H_B$ as an infinite sum (in the case
where the $w, z_{a}$ directions are compact) over modes.  This series
turns out to be absolutely convergent, and so is a useful
representation of the full solution.

We proceed here in the same way.  As stated above, we suppose that the
coordinates $z_{a}$, $w$ label compact tori of volume $V_a, V_w$.  If
these directions are not in fact compact, then a similar argument
follows simply by replacing the mode sums with integrals.  We Fourier
transform over the $z_{a}$, $w$ directions and decompose the solution
into harmonics on the $d-1$ sphere associated with the $d$ transverse
coordinates $x_\perp$.

In fact, it is only necessary to consider modes which are constant
over the $S^{d-1}$ sphere.  The higher spherical harmonics would tell
us about localization of the type B branes in the angular directions,
but we already expect that, when we move the type B branes to $r_\perp
= 0$, the limiting solution becomes symmetric over the $S^{d-1}$
sphere.  Including a discussion of the higher spherical harmonics, as
was done in \cite{SM}, verifies this conclusion, but does not impact
the question of localization in the $z_{a}$ and $w$ directions.  Thus,
we consider here only modes which are uniform over the $S^{d-1}$
sphere.  This is equivalent to replacing the point source in $B$ with
an $S^{d-1}$ shell of source: $q_B r_\perp^{-(d-1)}
\delta (r_\perp, r_{\perp 0}) \delta(w,w_0)
\delta (z_{a}, z_{a 0})$, where $r_{\perp 0} = |x_{\perp 0}|$.

We express the function $H_B(r_\perp, w, z_{a})$ as a sum over
Fourier modes in $w$ and $z_{a}$ multiplied by
radial modes $H_{B,p_w,p_{a}}(r_\perp)$, where $p_w$ and $p_{a}$ denote
the relevant (discrete) momenta in the $w$, $z_{a}$ directions.  These
radial modes satisfy the following second order ODE:
\begin{eqnarray} \label{rad} r_\perp^{-(d-1)} \partial_{r_\perp}
\left( r_\perp^{(d-1)} \partial_{r_\perp}
H_{B,p_w,p_{a}} (r_\perp) \right) -
p_w^2 H_{B,p_w,p_{a}} (r_\perp) & & \cr -
H_A (r_\perp) p_{a}^2 H_{B,p_w,p_a} (r_\perp) &=&
{{q_B} \over {V_a V_w}} r_\perp^{-(d-1)} \delta (r_\perp, r_{\perp 0}).
\end{eqnarray}

Let us, for the moment, fix our attention on one choice of $p_w,
p_{a}$, so that we need not indicate these labels explicitly.  The
$p_w=0, p_a=0$ mode has special boundary conditions which we will
discuss later.  For the other modes, imposing boundedness at large
$r_\perp$ will determine the solution at $r_\perp > r_{\perp 0}$ to be
a constant $\alpha_+$ times some particular solution $\phi_+$, while
continuity at the origin will determine the solution for $r_\perp <
r_{\perp 0}$ to be a constant $\alpha_-$ times some particular
solution $\phi_-$.  The constants $\alpha_\pm$ are then determined by
the matching conditions dictated by the delta-function in (\ref{rad}).
Namely, as the source in (\ref{rad}) has no derivatives of delta
functions, we must have $\alpha_+ \phi_+ (r_{0 \perp}) =\alpha_-
\phi_- (r_{0 \perp})$, while the discontinuity in the first
derivatives must reproduce the delta function source.  Using
$\phi_\pm$ and $\phi_{\pm}{}'$ to denote the values of the solutions
and their $r_\perp$ derivatives evaluated at $r_\perp = r_{\perp 0}$,
we have as usual

\begin{equation}
\label{coeff}
\alpha_+ = {{ q_B r^{-(d-1)}_{\perp 0}
 \phi_-} \over {V_a V_w(\phi_-\phi_+{}' - \phi_+ \phi_-{}')}},
 \ \ \
\alpha_- = {{q_B r^{-(d-1)}_{\perp 0}
\phi_+} \over {V_aV_w(\phi_-\phi_+{}' - \phi_+ \phi_-{}')}}.
\end{equation}
Our main task is to study the behavior of $\alpha_+$ as $r_{\perp 0}
\rightarrow 0$.  If the coefficient $\alpha_+$ vanishes in this limit,
then the corresponding mode will not appear in the limiting solution
for $H_B$.

The denominator of (\ref{coeff}) is just the Wronskian ($W$) of
(\ref{rad}) evaluated at $r_\perp = r_{0 \perp}$ and may be computed
by standard techniques.  Up to a constant (which does not depend on
$r_{\perp 0}$), it is given by $W(r_\perp)= r_\perp^{-(d-1)}$.  Thus,
up to an irrelevant constant we have
\begin{equation}
\label{co}
\alpha_+ = {{q_B
 \phi_- (r_{\perp 0})} \over {V_a V_w}},
 \ \ \
\alpha_- = {{q_B
\phi_+ (r_{\perp 0})} \over {V_a V_w}}. \end{equation}
All that remains is to determine the behavior of $\phi_-$ for small
$r_{\perp 0}$.

Let us suppose that $p_{a}$ is nonzero, so that the behavior near
$r_{\perp}=0$ is controlled by the term proportional to $H_A(r_\perp)
p_{a}^2$ in the radial equation (\ref{rad}).  Recall that $H_A$
describes some power law potential in $r_\perp$ that diverges at
$r_\perp = 0$.  For the $r_\perp^{-2}$ potential ($d=4$), the radial
equation may be solved exactly in terms of Bessel functions as was
done in \cite{SM}.  Given the explicit solution in \cite{SM}, one can
see that the sum over $p_{a}$ converges absolutely at any $r_\perp
\neq r_{\perp 0}$ so that the series gives an accurate description of
the physics.  Solutions to the source-free equation behave like
$r_\perp^{-1 \pm\sqrt{1+p_a^2 \hat{r}_A^2}}$ near $r_\perp =0$, so
that continuity at the origin forces $\phi_-$ to vanish there.  Thus,
only modes with $p_{a} = 0$ contribute in the $r_{\perp 0} \rightarrow
0$ limit and the type B branes delocalize in the $z_{a}$ directions.
This case corresponds to, for example, D1 branes parallel to D5
branes.

Still considering the case $d=4$ (e.g., D$1$-branes and D$5$-branes),
it is interesting to ask about the rate at which the type B branes
delocalize.  Since we are interested in the limiting behavior as $r
\rightarrow 0$, we consider the delocalization as it reaches the largest
length scales (small $p_{a}$).  Note that, as measured by the
asymptotic fields, the solution has delocalized on a length scale
$\delta x$ when the coefficient $\alpha_+$ corresponding to the
momentum $p_{a}\sim 1/\delta x$ becomes small relative to its value at
large $r_{\perp 0}$.  For convenience, we will measure $\alpha_-$
relative to its value ($ q_B\phi_-(\hat r_A)/V_aV_w$) at $r_{\perp 0}
= \hat r_A$.  A comparison with \cite{SM} or an approximate solution
for small $r_\perp$ shows that $\alpha_+$ vanishes like
\begin{equation}
{{V_a V_w \alpha_+} \over { q_B
\phi_-(\hat r_A)}} \sim \left( {{r_{\perp 0}} \over {\hat
r_A}}
\right)^{-1 + \sqrt{1+  p_{a}^2 \hat r_A^2}}
\end{equation}
for $r_{\perp } \ll \hat r_A$.  Specifically, for small $p_{a}$,
$\alpha_+$ vanishes like $\left( r_{\perp 0} / \hat r_A
\right)^{ {1 \over 2} p_{a}^2 \hat r_A^2}$.
Thus, for $d=4$, the type B brane appears to be delocalized on a
length scale $\delta x \sim \hat r_A \sqrt{ \ln {\hat r_{A} \over
{r_{\perp 0}}}}$ as viewed from infinity.  Note that, for large $\hat
r_A$, the B-branes are quite well delocalized before they reach any
strong curvature region.

Let us now consider the case $d>4$, e.g. D$0$-branes approaching
D$4$-branes.  Since $\phi_-$ vanishes at the origin for the $r^{-2}$
potential, one may expect the same behavior for the stronger
potentials $r^{-(d-2)}$.  This may be verified by looking for a
solution of the form $\phi = r_\perp^{-(d-1)/2} e^{\Psi (r_\perp)}$,
using the WKB approximation, and again imposing continuity at $r_\perp
=0$.  Note that the WKB approximation is self-consistent for such
strong potentials.  Thus, the supergravity solutions delocalize in the
$z_{a}$ directions for these cases as well.  The delocalization is
even faster than for $d=4$ as $\alpha_+$ now vanishes like
\begin{equation}
\alpha_+ \sim
\exp\left[ - p_{a} \left( {{d-2} \over 2} -1 \right)^{-1}
\hat r_A \left( {{\hat r_A} \over {r_{\perp 0}}}
\right) ^{{{d-2} \over 2} - 1} \right],
\end{equation}
as opposed to the power law behavior for the $r_\perp^{-2}$ potential.
When the type B branes are at $r_{\perp 0}$, the coefficients
$\alpha_+$ are small for $\hat r_A p_{a} > (r_{\perp 0}/\hat
r_{A})^{{{d-2}
\over 2} -1}$ and the type B brane has delocalized to a size scale
$\delta x \sim \hat r_A (\hat r_A/r_{\perp 0})^{{{d-2} \over 2} -1}$.
Again, at least within the domain of the WKB approximation, the sum
over momenta converges absolutely.  The case $d>4$ includes D0 branes
approaching D4 branes, as well as intersecting brane solutions (either
R-R, NS-NS, or M) with 5 or more directions transverse to both branes.
For example, it addresses the case of a D3-brane (B) and an orthogonal
D1-brane (A) smeared in the three $z_b$ directions along the D3-brane.
As the transverse separation is removed, the D3 brane delocalizes
along the D1-brane.  The case $d> 4$ also includes the case of
D-instantons in a D3 brane, which was explicitly seen to delocalize in
section \ref{intro}. As before, for large $\hat r_A$, the type B
branes are well delocalized before they reach any strong curvature
region.

The remaining case is when the bigger brane has only three transverse
directions ($d=3$), which is exactly the situation that arises in the
solutions of \cite{ITY}.  In this case, the WKB approximation is not
self-consistent, but we may study the radial equation by a related
technique.  Writing $\phi (r_\perp) = r^{-1} e^{\Psi (r_\perp)}$, the
sourceless radial equation (\ref{rad}) becomes

\begin{equation}
\partial_{r_\perp}^2 \Psi + (\partial_{r_\perp} \Psi)^2 - p_w^2 -
p_{a}^2 {{\hat r_A} \over {r_\perp}} = 0.
\end{equation}
This may be analyzed by assuming that $\partial^2_{r_\perp} \Psi$ is
much larger than $( \partial_{r_\perp} \Psi)^2$, which turns out to be
self-consistent for $r_\perp \ll \hat r_A$.  Within this
approximation, the general solution behaves near the origin like
\begin{equation}
\phi = C_1 \left({{r_\perp} \over {\hat r_A}}\right)^{[p_{a}^2 \hat
r_A r_\perp
-1]} + C_2 
\left({{r_\perp} \over {\hat r_A}}\right)^{[p_{a}^2 \hat r_A r_\perp]}.
\end{equation}
Thus, if $\phi$ is to be continuous at the origin we must have $C_1=
0$.  As a result, $\phi_-$ behaves like $\left( r_\perp / \hat
r_A\right)^{[p_{a}^2\hat r_A r_\perp]}$, which is finite and nonzero
at $r_\perp =0$.  This time we find that $\alpha_+$ does not vanish as
$r_{\perp 0}\rightarrow 0$.  The result is that the type B brane
remains localized in the $z_{a}$ directions in agreement with
\cite{ITY}.  Again, within the domain of validity of this
approximation ($ \hat r_A
\gg r_{\perp 0}, r_\perp$), one finds
that the sum over modes converges absolutely.  The infinite series
discussed here should sum to the solution of \cite{ITY} in the
near-core region.

We have not yet addressed localization in the $w$ directions for any
value of $d$.  Let us therefore consider a mode with $p_{a} = 0$, $p_w
\neq 0$.  The radial equations in this case are just those for the
Coulomb potential of a massive field and are easily studied.  One
finds that $\phi_-$ does not vanish at $r_\perp =0$, and the sum over
modes once again converges absolutely\footnote{Of course, the full sum
over modes includes a sum over $p_w$ even for $p_{a}\neq 0$.  For
$p^2_y \gg p_{a}^2 \left( {{\hat r_A} \over {r_\perp}} \right)^{d-2}$,
the analysis is identical to the $p_{a} = 0$ case and shows
convergence of the sum over $p_w$.}.  Thus, localization is always
possible in any directions transverse to brane B along which brane A
has been smeared.

The last mode to consider is the case $p_{a}=0, p_w=0$; i.e., the
spatially homogeneous mode.  This mode does not tell us about
delocalization of the branes; instead, it is the entire field
remaining once complete delocalization has occurred.  This mode is
special as its boundary condition at infinity differs from that of any
other mode.  The point is that, for $q_B=0$, the appropriate solution
is $H_B=1$ and not $H_B=0$.  Thus, the correct solution for the
homogeneous mode is a constant plus a function that vanishes for large
$r_\perp$.  We now consider the case where the type B brane
delocalizes completely and make several observations.  Recall that
complete delocalization occurs when $d \ge 4$ and there are no $w$
directions.  First, note that the complete solution for $r_{\perp 0}
=0$ is of the form $H_B = 1 + (r_B^{d-2+a}/V_a r_\perp^{d-2}),$ since
only the spatially homogeneous mode survives.  Second, since all modes
with $p_a \neq 0$ vanish at $r_{\perp}=0$ for any location $r_{\perp
0}$ of the type B branes, we may evaluate $H_B$ exactly at the origin:
\begin{equation}
\label{origin}
H_B(r_\perp =0) = 1 + \left({{ r_B^{d-2+a}} \over {V_a
r^{d-2}_{\perp 0}}}  \right).
\end{equation}
Note that this result also holds in the infinite volume limit ($V_a
\rightarrow \infty$) in which $r_B^{d-2+a}/V_a =0$.  Thus, if the $z_a$
directions are not compactified, we have simply $H_B =1$ at $r_\perp =
0$, independent of $r_{\perp 0}$.  These observations will prove
useful in the following sections.

We now make a few final remarks about variations on the above theme.
Consider, for example, solutions representing not branes in
asymptotically flat space, but branes in the near horizon geometry
associated with the type A branes; i.e., the solutions obtained by
taking a limit $\hat r_A \gg r_\perp, r_{\perp 0}$.  This is really
the only part of the asymptotically flat geometry of which we have
made significant use, so the discussion is not changed.  Note that
taking this limit is equivalent to setting
$H_A=({{\hat{r}_A}\over{r_\perp}})^{d-2}$ and fixing the boundary
conditions by imposing boundedness at infinity and specifying the
value of $H_B$ at $r_\perp =0$ to be given by (\ref{origin}) above.
The point we wish to emphasize is that the $p_a=0, p_w=0$ mode still
approaches a nonzero constant far from the type B branes and we have
$H_B\rightarrow 1$ as $r \rightarrow \infty$. One can see this
explicitly in the D-instanton example from the introduction, which
already resides in the near-horizon geometry of the three-brane.

In addition, one might ask how the analysis would change if a given
set of (type IIA) branes were lifted to eleven dimensional
supergravity.  Such a lift has a translational symmetry in $x_{11}$,
so that $x_{11}$ does not become a transverse direction.  As the
behavior of the classical solution is determined by the number of
dimensions transverse to the type A brane, it is not affected by this
process. One might also ask about related solutions where the M-brane
is localized in $x_{11}$.  An example where this is possible is
lifting a solution with D2-branes (with one brane smeared over the
other) intersecting in points to a solution with M2-branes
intersecting in points (with one brane still smeared over the other
but otherwise fully localized).  In such cases, the M-theoretic
solution will generally delocalize faster than the type IIA
supergravity solution due to the increased number of transverse
dimensions.  Note that a qualitative change from delocalization to
localization could only happen if the number of transverse directions
crosses the threshold at $d=4$, i.e., the lifting of a D6 brane to
M-theory.  But this lift is a Kaluza-Klein monopole and, while there
are only 6 directions along the corresponding M-theoretic `brane,' the
Kaluza-Klein monopole has a nontrivial structure in the remaining 4
spatial directions.  Due to the structure of these dimensions, the
corresponding potential $H_A$ behaves only as $r^{-1}$ even for the
M-theoretic solution, and therefore has the same properties with
regard to delocalization as the IIA D6-brane.  One can, of course,
consider a Kaluza-Klein monopole in any dimension and, as it always
has an $r^{-1}$ potential, the type B branes always remain localized.
This is of course to be expected from the method of
\cite{ITY,aki}, which realizes the `near-core' Kaluza-Klein monopole
as an orbifold of flat space and inserts the type B branes before
taking the orbifold quotient.

Finally, within supergravity, it is clear that the above analysis can
be extended to branes that are not asymptotically flat.  They
correspond to potentials that are softer than $r^{-1}$ and therefore
allow localization of the directions parallel to the A brane.
However, this case is not our primary concern and we will not discuss
it in detail.

\section{The supergravity / field theory correspondence.}
\label{IFT}

In the last section we studied the delocalization behavior of a class
of asymptotically flat supergravity brane solutions of type IIA/B
string theory.  We would now like to understand this behavior from a
field theory perspective by using a generalization of the $p$-brane
AdS/CFT correspondence, which is obtained by taking a certain
low-energy limit of a system of $N_p$ R-R charged $p$-branes.  We
first study the supergravity solutions and define the AdS/CFT limit,
and then move on to a field theory explanation of the supergravity
(de)localization phenomena.

Let us first orient ourselves with a lightning review of the salient
features of the AdS/CFT correspondence of \cite{IMSY} for D$A$-branes.
Here, $A$ denotes the number of spatial dimensions of what in section
\ref{SUGRAV} was a type A (Dirichlet) brane.  We will ignore numerical
factors in this entire section; the relevant precise normalizations
may be found in \cite{IMSY} or in the previous section.  The AdS/CFT
correspondence for $N_A$ D$A$-branes is obtained by starting in string
theory with the coupled bulk-brane dynamics, and taking the low-energy
limit
\begin{equation}\label{lowen}
(E \ell_s) \rightarrow 0 \,.
\end{equation}
This energy $E$ is measured in the gauge theory.  The gauge theory
coordinates are the isotropic supergravity coordinates in directions
parallel to the brane.  To see this, one can compute the associated
moduli space metric for motion of a probe D$A$-brane in the background
of the others; it is trivial.  The next step is to keep the
dimensionless expansion parameter in the U($N_A$) gauge theory fixed;
since we are interested in large numbers of branes for comparison to
the supergravity, this expansion parameter is
\begin{equation}
\label{lambdadef}
\lambda_A^2(E) = g_{YM,A}^2 N_A E^{A-3} = g_s N_A (\ell_s E)^{A-3} \,.
\end{equation}
The field theory is perturbative when this parameter is small.  One
also keeps fixed the mass of strings stretched perpendicular to the
branes, $E=|x_\perp|/\ell_s^2\equiv U$, where $x_\perp$ is the
displacement between the branes.

In the AdS/CFT low-energy limit the physics on the brane decouples
from the physics in the bulk, provided that $A<6$.  This decoupling
happens essentially because supergravity is a weak interaction at low
energy compared to the gauge interaction.

The supergravity metric is
\begin{equation} \label{dsa2}
ds^2 = {{1}\over{\sqrt{H_A}}}\left(-dt^2 + dz_A^2\right) +
\sqrt{H_A}dx_\perp^2 \,, \qquad {\rm where\ \ }
H_A=1+\left({{r_A}\over{|x_\perp|}}\right)^{7-A} \,.
\end{equation}
Here, we have taken all of the D$A$-branes to be `clumped' together at
the origin $r=0$ and have not allowed any smearing of the branes.  The
symbol $r_A$ denotes the charge radius of the brane, i.e. the radius
where the 1 in the harmonic function is comparable to the other term:
\begin{equation} \label{rAdef}
r_A = \ell_s (g_sN_A)^{1/(7-A)} \,.
\end{equation}
Converting $x_\perp$ to $U$ and the other factors using
(\ref{lambdadef}) the harmonic function may be written
$H_A = 1 + \lambda^2_A(U)/(\ell_s U)^4$.
Assuming that the supergravity $U$, which is an energy, scales the
same as the gauge theory energy $E$, we see that the 1 in $H_A$ is
lost in the low-energy limit.  This means that we have lost the
asymptotically flat part of the supergravity geometry.  In considering
the physical validity of this near-horizon supergravity solution,
there are two types of corrections to worry about, $\alpha^\prime$ and
$g_s$.  A measure of the first type of correction is the Ricci scalar
measured in string units, which is found to be
$\ell_s^2 {\cal{R}} = 1/\lambda_A(U)$
and so the supergravity is weakly coupled for $\lambda_A(U)\gg 1$.  In
comparing this to the gauge theory regime $\lambda_A(E)\ll 1$, as
pointed out in \cite{JGPAWP}, we must be careful in specifying the
type of probe we are using, which in turn gives a relationship between
the gauge theory energy $E$ and the supergravity radius $U$.  The
simplest type of probe to consider is the one originally studied in
\cite{IMSY}, the stretched fundamental string, which is a BPS state in
the gauge theory and has $E=U$ as above.  One then sees that the
supergravity and gauge theory breakdowns happen in a consistent
fashion.  Here it was important that this result for the stretched
string mass is not corrected gravitationally.  The gravitational
`warpage' for a fundamental string is computed from the D$A$-brane
metric in the string frame (\ref{dsa2}), and since the (BPS) stretched
string points in a direction perpendicular ($\perp$) to the branes the
warpage is unity:
\begin{equation}
\perp {\rm warpage\ } = \int \sqrt{-g_{\tau\tau}g_{\sigma\sigma}}
= \int \sqrt{H_A^{-1/2}H_A^{+1/2}} =1 \,.
\end{equation}
At still higher values of $\lambda$, other descriptions of the physics
are appropriate \cite{IMSY}, such as S-dual supergravity geometries.
Overall one finds different descriptions of the physics to be valid at
different $\lambda^2$'s, and they are dual to one another.  We will be
concentrating on the super Yang-Mills and ten-dimensional supergravity
phases.

Lastly, let us make a brief remark contrasting the near-horizon
spacetime with the asymptotically flat solution.  If we compute the
curvature in string units of the asymptotically flat $A<3$ solutions,
we find that it starts from zero at $x_\perp=0$, peaks at around the
charge radius at a value of order $(g_s N_A)^{-2/(7-p)}$ which is
small for large $g_sN_A$, and falls off to zero again at
$r\rightarrow\infty$.  For $A>3$, a similar behavior occurs for the
dilaton. This is to be contrasted with the near-horizon spacetime for
which the curvature and dilaton are monotonic \cite{IMSY}.  For $A<
3$, the curvature diverges near the branes while for $A>3$ the dilaton
diverges there.

\subsection{The supergravity side}

We now wish to define a generalized AdS/CFT correspondence for
intersecting D-branes appropriate to the R-R supergravity solutions
that we constructed in the last section. Our conventions differ from
the last section only in that, for the moment, we take there to be no
$w$ directions.  One special case in which we will need to reintroduce
$w$ directions will be discussed near the end of subsection B.  So,
our setup is
\begin{center}
\begin{tabular}{|c||c|c|c|c|c|} \hline
{}   \ & \  $t$       \ & \  $z_I$     \ & \  $z_a$     \ & \  $z_b$
  \
& \  $x_\perp$
        \\ \hline
D$A$ \ & \  $\bullet$ \ & \  $\bullet$ \ & \  $\bullet$ \ & \
$\equiv$  \
& \  {}
        \\ \hline
D$B$ \ & \  $\bullet$ \ & \  $\bullet$ \ & \  {}        \ & \
$\bullet$ \
& \  {}
        \\ \hline
\end{tabular}
\end{center}

\noindent There are $d=(9+I-A-B)$ of the $x_\perp$,
$a=(A-I)$ of the $z_a$, and $b=(B-I)$ of the $z_b$.  Without loss of
generality, we have chosen the D$A$-branes to be smeared over the
$z_b$.  
If there are no $z_b$, then
the intersection is parallel, otherwise it is orthogonal. In addition,
we take any smeared configuration to be irreducibly smeared in the
sense that it cannot be reduced via a T-duality symmetry
transformation to an unsmeared configuration.
We work in the large-volume limit, $V_{a,b}\rightarrow\infty$,
but we keep a finite charge density of the A-branes,
$N_A/(V_b/\ell_s^b)$.  In contrast, the charge density
$N_B/(V_a/\ell_s^a)\rightarrow 0$.    Our analysis therefore
differs from the special case
considered in \cite{TL} in which there was a volume
infinity of D-instantons on D3-branes\footnote{For this case, in
contrast to what we study later, the theory on the D-instantons is
not dynamical.}.  It also differs from previous analyses such
as \cite{BooPS} where AdS/CFT for smeared intersecting branes
was studied.  

Let us now study our intersecting brane supergravity solutions from
section \ref{SUGRAV} in a generalization of the low-energy limit of
\cite{IMSY}.  Our concern is to find the relevant region of the
supergravity geometry.  If the D$A$-branes and D$B$-branes are very
far from each other, then each has its own near-horizon region as if
the other collection of branes did not exist.  The more interesting,
and new, situation arises when the branes influence each other
strongly, i.e. when the separation $x_{\perp 0}$ is near-horizon in a
sense which we now explain.  In this section, we keep the localization
in the $(d-1)$-sphere explicit, and do not use spherical shells.  

An important subtlety in determining the low-energy limit for
intersecting D-branes concerns the kind of probe we will use to
connect a gauge theory energy $E$ with a supergravity radius $U$.  As
before, the most straightforward probes to consider are the stretched
strings.  In the intersecting brane geometry, with metric
(\ref{intbraneg}), and `harmonic' functions
$H_A(x_\perp),H_B(x_\perp,z_a)$, the only BPS stretched strings are
the ones running in the transverse ($x_\perp$) directions.  Strings
stretched in directions other than $x_\perp$, for example tethering
two clumps of B branes separated in a $z_a$ ($\parallel$) direction,
are not BPS; a symptom of this is that they experience large warpage
\begin{equation}
\parallel {\rm warpage\ } = \int \sqrt{-g_{\tau\tau}g_{\sigma\sigma}} 
= \int d\tau d\sigma \sqrt{1/H_A} \,.
\end{equation}
Because of this warpage, the non-transverse strings are attracted to
the A branes.  So we will concentrate only on the transverse BPS
stretched strings in the following; the dynamics for other types of
probes is much more complicated and we will not discuss it.  As a
consequence, when we take the low-energy limit, we perform the scaling
of coordinates only in the transverse directions:
$U\ell_s=x_\perp/\ell_s\rightarrow 0$.

If $A\geq B$ we hold the A-brane coupling $\lambda^2_A(U)$ fixed; if
$A<B$ we hold $\lambda^2_B(U)$ fixed.  Effectively, this means we are
taking the AdS/CFT limit just as for the larger brane.  In a sense,
including the physics of the smaller brane is like a `perturbation' on
this larger-brane AdS/CFT correspondence, although we are not in any
way treating the smaller brane physics perturbatively.  A simple
example of this is the case of the D-instanton in the D3-brane
near-horizon geometry which we discussed in the introduction; however,
for the examples with larger branes such as D1$\parallel$D5, the
theory on the smaller branes is in fact dynamical.  In addition, the
intersecting brane gauge theories will decouple from the bulk as long
as the branes are `small enough'; here this means that there are more
than three directions transverse to both branes.  For the cases where
there are only three, even though the metric does not have the same
form as for a D6-brane, we can compute the curvature and Hawking
temperature and they turn out to behave as for sixbranes.  Therefore
we suspect that holography is breaking down, i.e. that making
predictions about the classical supergravity from the quantum brane
field theory may be problematic.  We will remark on these cases
explicitly when we encounter them.

Now, doing the AdS/CFT scaling in $x_\perp$ and holding $\lambda^2$ of
the larger brane fixed gives the the near-horizon geometry of the
D$A$-branes.  That this is true no matter whether $A
\ge B$ or $A<B$ can be seen as follows.  {}From
\begin{equation}
H_A=1+ {{r_A^{7-A}}\over{V_b}}{{1}\over{|x_\perp|^{7-A-b}}}
\end{equation}
we find
\begin{equation}
\label{aha}
H_A  \rightarrow
{{\left[g_s N_A (\ell_s U)^{A-3}\right]}
        \over{(V_b/\ell_s^b)(\ell_s U)^{4-b}}}
= {{N_A}\over{(V_b/\ell_s^b)N_B}}
{{\left[g_s N_B (\ell_s U)^{B-3}\right]}\over{(\ell_s U)^{4-a}}}.
\end{equation}
Recall that there are several constraints of our setup, such as
asymptotic flatness of the intersecting brane solution, the smearing
of the D$A$-branes over $z_b$, and the fact that both intersecting
D-branes live in either Type IIA or IIB string theory.  As a result,
it is always true that $b<4$ if $A\geq B$ where we hold
$\lambda^2_A(U)$ fixed, or $a<4$ if $A<B$ where we hold
$\lambda^2_B(U)$ fixed.  Thus we see that, as previously advertised,
in either case only the near-horizon part of $H_A$ remains.  That this
happens even for $A<B$ is due to the smearing of the A-branes over
$z_b$.

We now need to check whether the supergravity curvature and dilaton
develop any additional strong coupling regions due to the presence of
the B branes.  First of all, we should not be tempted to just throw
away the 1 in $H_B$ by analogy with what happened to $H_A$.  Recall
that in the infinite volume limit, we have $H_B =1$ at the location of
the D$A$-branes for any value of $x_{\perp 0}$, and that this provides
a boundary condition which forces $H_B\rightarrow 1$ far from the
D$B$-branes as well.  (We also discussed this phenomenon explicitly
for the $A=3,B=-1$ example in the introduction.)  We conclude that,
far from the D$B$-branes, the geometry and dilaton are just as they
were for the near-horizon D$A$-branes in isolation.  Thus, one expects
that new strong coupling regions could arise only near the
D$B$-branes.  We examine this possibility now, though we need to treat
the $x_{\perp 0}=0$ and $|x_{\perp 0}|>0$ cases separately.  Let us
study $x_{\perp 0}=0$ first.

For $x_{\perp 0}=0$ we go directly to the exact solution of the
previous section.  For all cases with potentials stronger than $1/r$,
where we expect holography to hold, delocalization occurs, and the
solution depends only on the $d$ overall transverse coordinates.  For
$x_{\perp 0}=0$ the solution is
\begin{equation}
H_B = 1 + {{r_B^{7-B}}\over{V_a}}{{1}\over{r^{7-B-a}}}
= 1+ {{g_s N_B}\over{V_a/\ell_s^a}}
\left({{\ell_s}\over{|x_\perp|}}\right)^{7-A-b}  \,.
\end{equation}
where $V_a$ is the volume of the $z_a$ directions, and we have used
the identity $a+B=A+b$.  Since we do not have a ($V_a$) volume
infinity of D$B$-branes, we have just $H_B=1$, everywhere.  As a
result, the curvature and dilaton behave exactly as they would for the
isolated D$A$-branes.  We can therefore simply use our D$A$-brane
intuition to tell us where the $x_{\perp 0}=0$ intersecting brane
supergravity solution is valid.

At generic separations $x_{\perp 0}$, the supergravity solution is
complicated, and is not known in terms of elementary functions. We
will instead use an approximation scheme to study the curvature.  As
usual, we consider D$A$-branes with worldvolume $\{t,z_I,z_a\}$ and
D$B$-branes with $\{t,z_I,z_B\}$.  The supergravity equation of motion
for $H_A$ is
\begin{equation}
\left[ \partial_{{x_\perp}}^2 \right] H_A = {{q_A}\over {V_b}}
\delta({{x}}_\perp)
\end{equation}
i.e. the $A$-branes are all at ${{x}}_\perp =0$.  The near-horizon
solution is
\begin{equation}
H_A = {{r_{A}^{7-A}}\over{V_b|{{x}}_\perp|^{7-A-b}}} \,,
\end{equation}
where the $b$ appears because of the smearing of the D$A$-branes along
the D$B$-branes.  (Recall that, in our low-energy limit, the 1 in
$H_A$ has dropped out.)

We chose the origin of coordinates such that the D$B$-branes are
located at ${{z}}_a ={{0}}$ and at ${{x}}_\perp={{x}_{\perp 0}}$.  The
equation satisfied by $H_B$ is
\begin{equation}
\label{eqws}
\left[ \partial^2_{{{x}}_\perp} + H_A \partial^2_{{{z}}_a}
\right] H_B = q_B
\delta({{x}}_\perp,{{x}_{\perp 0}}) \delta({{z}}_a) \,.
\end{equation}
We will solve this equation in the region near the D$B$-branes, in
particular for ${x}_\perp$ satisfying
$|{{x}}_\perp-{{x}_{\perp 0}}|\ll|{{x}_{\perp 0}}|$.
Over such a region, the function $H_A$ does not vary much, and we can
approximate it by a constant in the equation for $H_B$.  After a
change of coordinates,
\begin{equation}
{{y}}_\perp = {{x}}_\perp-{{x}_{\perp 0}} \,, \qquad
{{y}}_a = {{{{z}}_a}\over{\sqrt{H_A}}} \,,
\end{equation}
we see that an approximate solution to the $H_B$ equation of motion is
given by
\begin{equation}
H_B = f_0 + {{r_B^{7-B}}\over{
\left[({{{y}}_a})^2+({{{y}}_\perp})^2\right]^{(7-B)/2}   }}\, ,
\end{equation}
where $f_0$ is a solution of the $q_B=0$ version of (\ref{eqws}). In
particular, $f_0$ is smooth at the location of the D$B$-branes, so
that the singular term will dominate in this region.  For convenience,
we pretend that $f_0 =1$ in order to borrow results from the familiar
D$B$-brane metric, but our conclusions will not depend on this choice.

Now notice that the fields of the intersecting brane system do not
depend on any of $\{t,{{z}}_I,{{z}}_b\}$.  We may therefore rescale
these coordinates without affecting the supergravity physics.
Defining
\begin{equation}
  (T,{{y}}_I,{{y}}_b) =
{{(t,{{z}}_I,{{z}}_b)}\over{\sqrt{H_A}}} \,,
\end{equation}
the metric becomes
\begin{equation} ds^2 = \sqrt{H_A}\left[
{{1}\over{\sqrt{H_B(y_a,y_\perp)}}}
\left(-dT^2+d{{y}}_I^2+d{{y}}_b^2\right)+
{\sqrt{H_B(y_a,y_\perp)}}\left(d{{y}}_a^2+d{{y}}_\perp^2\right)
\right]\,.
\end{equation}
With the metric near the D$B$-branes in the above form, the scalar
curvature ${\cal{R}}$ of the intersecting brane solution is easy to
compute.  We have
\begin{equation}
\ell_s^2 {\cal{R}} \sim {{1}\over{\sqrt{H_A}}}\ell_s^2 {\cal{R}}_B
\end{equation}
where ${\cal{R}}_B$ is the curvature of the D$B$-brane metric by
itself.  At large $g_s N_B$, this curvature is small everywhere, as we
saw at the beginning of this section.  The factor of $1/\sqrt{H_A}$
only makes this conclusion stronger, by a factor
$\sqrt{(g_sN_A)/(x_{\perp 0}/\ell_s)^{7-A-b}}\rightarrow\infty$.

We also consider the dilaton.  For the intersecting solution it is
\begin{equation}
e^\Phi = g_s H_A^{(3-A)/4} H_B^{(3-B)/4} \,.
\end{equation}
For small $B$, this will become large near the D$B$-branes.  For $A>3
$, it will be damped by a power of $H_A$.  However, due to the change
of coordinates above from $ z_a$ to $ y_a$, $H_B$ is large over a
range of $ z_a$ that is larger than for the B-branes in isolation.
For small B, the supergravity solution also breaks down near the
D$B$-branes, over a range of ${z}_a$ that is significantly stretched
relative to what occurs for the B-branes by themselves.  If we were
interested in the region where the dilaton were too large, we would
switch to an S-dual description.

However, the most important point for us follows from the observation
of the previous section that we have $H_B\rightarrow 1$ far from the
D$B$-branes.  In fact, when the separation $| x_{\perp 0}|$ is small
and if we do not have a volume ($V_a$) infinity of D$B$-branes, $H_B$
approaches $1$ quite rapidly.  Now let us pick a point significantly
further out than the D$B$-branes but close enough that it would lie
inside the region of validity of the near-horizon supergravity
solution for the D$A$-branes alone.  Then in the combined system,
$H_B$ is close to 1 at this point, and the D$B$-branes do not affect
the validity of the supergravity solution.  The previous section then
tells us that the delocalization is visible in this region, and also
that, by adjusting parameters, we can make the delocalization
arbitrarily large without placing the D$B$-branes in the
strong-curvature region of the near-$A$-brane geometry.  We conclude
that delocalization is a reliable prediction in this intersecting
brane spacetime, and so it should have a dual description in terms of
the gauge field theory on the branes.

\subsection{The field theory side}

On the field theory side of our generalized correspondence we have the
coupled field theory of the D$A$- and D$B$-branes.
There are three sectors of open strings, the
$A\!-\!A$ strings, the $B\!-\!B$ strings, and the
$A\!-\!B$ strings.  The action for this system in the
low-energy limit at weak gauge couplings is well known; it is T-dual
to that for the D0$\parallel$D4 system and has eight real supercharges.

Since the dimensions of the low-energy field theories for the
$A\!-\!A$, $B\!-\!B$, and $A\!-\!B$ strings are all different, we need
to know how the couplings scale relatively in the low-energy limit
$(E\ell_s)\rightarrow 0$.  The gauge couplings on the $A$- and
$B$-clumps are built out of the same string theory parameters.  Let us
study the expansion parameters $\lambda^2$ closely.  For the $A$-clump
we have
\begin{equation}
\lambda_{A}^2(E) = g_s N_{A} (\ell_s E)^{A-3} \,,
\end{equation}
and for the $B$-clump
\begin{equation}
\lambda_{B}^2(E) = g_s N_{B} (\ell_s E)^{B-3} \,.
\end{equation}

Now let us take the low-energy limit of a system of two clumps of
branes, the D$A$-branes and the D$B$-branes, as we did in the previous
subsection.  In general the expansion parameters will develop a
hierarchy for energies $E$ of similar order:
\begin{equation}
{{\lambda^2_A(E)}\over{\lambda^2_B(E)}} \sim {{N_A}\over{N_B}} (\ell_s
E)^{A-B} \,.
\end{equation}
We see that a large hierarchy arises because we are taking the
low-energy limit $(\ell_s E)\rightarrow 0$.  Now recall the condition
that we keep $\lambda^2(E)$ fixed for the bigger brane; this would be
$\lambda^2_A(U)$ for $A\geq B$ or $\lambda^2_B(U)$ for $A<B$.
Therefore we see that the coupling of the smaller brane becomes much
stronger than that of the larger one,
\begin{equation}
\lambda^2_{\mbox{\scriptsize{small}}}(E) \gg
\lambda^2_{\mbox{\scriptsize{big}}}(E) \,.
\end{equation}
This means that the physics on the bigger brane does not significantly
influence the physics on the smaller brane.  This is similar in
spirit to the AdS/CFT decoupling of the bulk theory from the brane
theory.  So whenever we consider two intersecting branes, orthogonal
or parallel, we need only study the gauge theory on the smaller brane
and on the intersection, and we can ignore the physics on the bigger
brane as it is essentially frozen out.  The one exception occurs, of
course, when $A=B$, in which case the dynamics on both clumps of
branes is equally relevant.

Now, for D$p$-branes with $p<3$, we have from the formula for
$\lambda_p^2(E)$ that the perturbative SYM regime is the high-energy
or ultraviolet regime \cite{IMSY}.  So in a Wilsonian sense the SYM
description is the fundamental one.  Now, note that the dimension of
the theory on the intersection for our intersecting D-brane
configurations has an upper bound of $d=2+1$.  This happens because
there are not enough dimensions of spacetime to have intersecting
branes (parallel or orthogonal) which are asymptotically flat and
which have an intersection theory with $d>2+1$.  Therefore, even at
strong coupling, we may rely on conclusions that follow from general
properties of the field theory on the smaller branes (with $p\le 3$)
and on the intersection, such as locality and dimensionality.  On the
other hand, the SYM physics on the $d=p+1$ worldvolume of $p>3$ branes
is at best a low-energy effective field theory, and gets replaced in
the UV by a more complicated theory which may not even be a local
quantum field theory \cite{IMSY}.  Taking into account our previous
finding that the gauge physics is relevant only on the intersection or
the smaller brane, or at worst on both clumps for the $A=B$ case, the
only case where the fact that SYM$_{p+1}$ is not the UV theory might
bother us is D4$\perp$D4(2).  In fact, since in this case we do not
know the fully localized supergravity solution but only an irreducibly
smeared one, we are in effect dealing with $1/r$ potentials.
Therefore, we suspect that holography may be breaking down for this
particular case.

We see that the strongly quantum mechanical coupled D$A$,D$B$ field
theory describes the supergravity solution in the near-$A$-horizon
regime.  This concludes our discussion of the regime of validity of
the gauge theory and its relation to the supergravity regime of
validity.  Now we turn to the gauge theory considerations, with which
we want to explain our previous supergravity results on
(de)localization.

In the supergravity section we saw (de)localization occur as we
brought the B-branes in to the A-branes from finite transverse
separation to zero separation.  We now want to see how this happens
from the field theory perspective.  Since the type A and B branes are
initially separated, we are a priori on the `separated' branch of
moduli space.  For the D5$\parallel$D1 (and T-dual D4$\parallel$D0)
case this is usually called the Coulomb branch, and the `coincident'
branch the Higgs branch.  Now, if we were to integrate out the 1-5
strings to study the Coulomb branch, we would find
\cite{Wit,ABKSM} that the Coulomb and Higgs branches of
moduli space are separated by an infinite distance and decouple.
However, these 1-5 strings become light in the limit in which the
separation between the A and B branes is removed, which is our
situation of interest.  In particular, in our setup we have kept the
mass of such strings fixed relative to our gauge theory energy scale.
Thus, the moduli space is simply not sufficient to describe physics in
the region of interest.  An analogy to our Ramond-Ramond case is the
S-dual situation of fundamental strings approaching N-S fivebranes.
There, the infinite distance in moduli space corresponds to the
infinite throat of the fivebrane. However, we know that this is no
obstacle to a fundamental string reaching and crossing the fivebrane
horizon.  Again, what one finds is that the moduli space approximation
is simply not sufficient to describe this part of the dynamics.  In
the same way, there is no conflict between our picture and the results
of \cite{Wit,ABKSM}; a mixing between the Coulomb and Higgs branches
is allowed in our setup of the AdS/CFT limit for the intersecting
branes when the separation is small\footnote{Our analysis also differs
in that our B-branes are not probes; we take into account their effect
on the supergravity fields.  In addition, the field theory description
of our setup is neither the conformal field theory which appears in
the extreme IR on the Higgs branch nor the one on the Coulomb branch
(these theories have different R-symmetries) \cite{Wit}.}.

Let us consider in more detail the case of D$1$-branes (B) and
D$5$-branes (A), where on the Higgs branch the relevant moduli are the
scale sizes and orientations of the gauge instantons which represent
the D1-branes in the D5-brane gauge theory.  (There are also position
moduli but they will not be important in the following.)  As we argued
above, we expect a mixing between the Coulomb and Higgs branches as
the separation goes to zero.  We will use physics of the Higgs branch
to study delocalization, in an approximate sense, keeping in mind that
the small mass of the 1-5 strings modifies the dynamics of the system
at large length scales and thus provides an IR cutoff.  Now let us
extract a length scale for instanton size fluctuations.  The scale
size and orientation for a single instanton form a 1+1 quantum field
theory with a moduli space metric which is flat for $N_1=1$, and with
a coefficient $1/(l_s^2 g_s)$ in front of the Lagrangian.  The $l^2_s$
is associated with the fact that $\rho$ has dimensions of a length.
That the metric is not renormalized at strong coupling is a
consequence of the high degree of supersymmetry in this system (it is
hyperK{\"a}hler).  For large $N_5$, there are roughly $N_5$ possible
orientations for the instanton in gauge space, so we have
\begin{equation}
\sqrt{  \langle \rho^2   \rangle} = \sqrt{ \langle
\rho_1^2 + \ . \ . \ . \ + \rho_{N_5}^2 \rangle }
\sim  \sqrt{N_5 g_s l_s^2} \sqrt {\log\left({\Lambda_{UV}/\Lambda_{IR}}\right)},
\end{equation}
where $\Lambda_{IR}, \Lambda_{UV}$ are appropriate infrared and
ultraviolet cutoffs.  Now, since we take all energies low by
comparison to the string scale as in (\ref{lowen}), $\Lambda_{UV} \sim
l_s^{-1}$.  In addition, as above, the 1-5 strings have a mass $U$
which provides an IR cutoff.  If there are $N_B$ separate instantons,
the moduli space metric, although uncorrected, is not flat.  What is
important for our estimate is the normalization, which is the same as
in the $N_B=1$ case.  In addition, the instantons all fluctuate
independently, and so we may expect the above rough estimate to carry
over.  To translate our estimate into the quantities used in the
classical supergravity discussion, recall that $U = r_{\perp 0}/l_s^2$
and, since the A-brane is an unsmeared five-brane, $r_5 = l_s
\sqrt{N_5 g_s}$.  Also, since we are holding 
$\lambda_5^2(U)=g_s N_5 (\ell_s U)^2$ fixed and $r_{\perp 0}$ is
small, up to numbers of order one we may replace the $l_s$ coming from
$\Lambda_{UV}$ inside the logarithm with $r_5$.  We then have
\begin{equation}
\sqrt{ \langle \rho^2 \rangle } \sim r_5 \sqrt{ \ln (r_{\perp 0} /r_5)}.
\end{equation}
We see that this estimate matches the supergravity result.

The story is similar for D$0$ branes approaching D$4$ branes.  In that
case, we have a 0+1 quantum field theory and the rms fluctuations will
be proportional to $\sqrt{1/\Lambda_{IR}}$, but $r_4 = (g_s
N_4)^{1/3} l_s$.  We find
\begin{equation}
\sqrt{\langle \rho^2 \rangle } \sim \sqrt{N_4 g_s l_s}
\sqrt{l^2_s/r_{\perp 0}} =
(g_s N_4)^{1/3} l_s \sqrt{{(g_s N_4)^{1/3} l_s}\over {r_{\perp 0}}}.
\end{equation}
Again, this matches the classical supergravity delocalization rate.
The case of D3-branes and D-instantons is a bit degenerate, but one
finds agreement with the classical supergravity results, and with
scale-radius duality, by taking
$\sqrt{\langle \rho^2 \rangle}$ proportional to
$1/\Lambda_{IR}$. 

We note that, for every case of intersecting Ramond-Ramond branes that
falls within our framework, we have $A+b \ge 4$ and the potential
$H_A$ diverges no faster than $r^{-3}$.  Furthermore, an $r^{-3}$
potential is always associated with a 0+1 dimensional intersection and
an $r^{-2}$ potential is always associated with a 1+1 dimensional
intersection.  Similarly, cases where $H_A$ diverges only like
$r^{-1}$ or weaker correspond to $2+1$-dimensional or larger
intersections, and both the quantum size moduli fluctuations in the
brane gauge theory and the classical supergravity delocalization are
small.  (If the $z_I$ were compactified on a very large manifold, the
above results hold in the infinite-volume limit, and so by continuity
the (de)localization results are essentially unchanged at large but
finite volume.  We will, however, avoid finite volumes so as to
finesse additional phenomena that occur when the sizes of the
compactified manifolds get too small near the core in the supergravity
geometry.)  Therefore, we see agreement for both the parallel and
orthogonal intersections.

In cases where $b > 0$ i.e.  the type A branes are smeared, their
field theory is still $A+1$ dimensional, not $A+b+1$ dimensional.  It
is therefore reasonable to replace $N_A$ in the argument above for the
delocalization rate with the volume density $N_A l_s^b/V_b$, as one
may think of this case as having a large number, of order $V_b/l_s^b$,
of separate intersections.  Thus, an estimate of the instanton scale
size fluctuations can always be made that agrees with the classical
supergravity delocalization rate.

Let us lastly consider the qualitatively different kinds of solutions
we get by additionally smearing the type A branes along the $w$
direction.  In the classical supergravity, we have a localized
solution if only three transverse dimensions are left unsmeared.  This
is hard to explain from the field theory perspective, as can be seen
by considering a prototypical example of the D0-clump with a D4-clump
smeared along two of the five transverse dimensions.  Then the smeared
D4-clump gives rise to a $1/r$ potential and the supergravity solution
localizes, even though the intersection is $0+1$ dimensional.  Now,
recall that holography for the D6-brane system is problematic.  For
the D4-brane smeared over two transverse directions, we have a $1/r$
potential and so we suspect that holography is breaking down for this
twice-smeared D4-brane as well.  In this sense, our success in getting
the classical supergravity answer for the D6$\parallel$D2(2) and
D4$\perp$D4(2) systems from the quantum brane gauge theory is
surprising.  On the other hand, it may simply be that smearing one
brane and not the other is not a straightforward operation from the
field theory perspective.  Note also that when the D$4$-branes are
smeared over only one transverse direction, instead of two, the
quantum theory on the branes and the classical supergravity agree that
delocalization should occur, but do not agree with regard to the rate
at which this happens.

\subsection{Asymptotically Flat Orthogonal Branes}

We would like to add a few more comments on the cases involving
orthogonal intersecting branes, and how to use our earlier results to
say something about asymptotically flat, as opposed to near-horizon,
spacetimes.  Recall that the intersection field theory description is
dual to the near-horizon supergravity description.  As such, it does
not directly say anything about the asymptotically flat solutions.
However, the near-horizon and asymptotically flat supergravity
solutions are controlled by the same equations of motion
(\ref{A},\ref{B}).  The only difference is in the boundary conditions
imposed on $H_A$; the boundary conditions on $H_B$ are identical for
both cases. Thus the two supergravity solutions must agree to high
accuracy in the region $|x_\perp| \ll r_A$. Thus, if the near-horizon
geometry is delocalized, there must also be a region (perhaps, only
for $|x_\perp| \ll r_A$) in which the asymptotically flat geometry is
delocalized as well.  We saw this explicitly for the solutions
exhibited in section \ref{SUGRAV}, in which the type A branes were
initially smeared over the $z_b$ directions.  The same conclusion
should hold in the case without the initial smearing, for which the
supergravity solutions are not yet known.  For initial progress toward
constructing these solutions, see \cite{l14.1}.

Let us now consider as a prototype of orthogonal intersections the
D2$\perp$D2(0) system. Initially, for clarity, we refrain from
smearing the A-branes over the B-brane world-volume.  In the quantum
gauge field theory, the instanton scale sizes become blowup modes of
the orthogonally intersecting D2-branes.  To see this, write each pair
of spatial worldvolume coordinates as a complex coordinate $Z$, then
for the combined worldvolumes we get the holomorphic curve $Z_A Z_B =
\rho$.  The smearing of $\rho$ is infinite because the field theory on
the intersection is only $0+1$ dimensional.  This means that the
corresponding near-horizon geometry will also be smeared.

We now use the above argument about matching supergravity solutions
and our knowledge of the blowup modes in the near-horizon case to draw
some conclusions about delocalization in the asymptotically flat case.
This solution should be delocalized, but perhaps only in some
near-horizon region.  A diagram giving our artistic impression of the
asymptotically flat solution is included below.  We can only conclude
that delocalization must occur in the interior of the shaded region,
which is the region inside a blowup mode that has expanded until it
reaches the curve $r=r_2$.  This is consistent with our expectations
that, far from the intersection, the solution should reduce to the
known physics of a lone D2-brane clump.  Our delocalization has become
a finite-sized `neck' of the supergravity solution.

\vskip-0.0truein\hskip1.0truein{\epsfbox{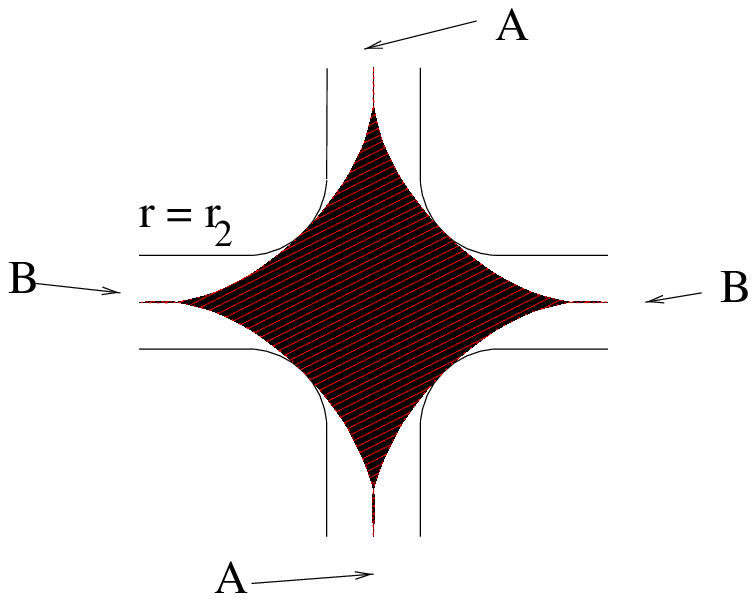}}

In the actual D2$\perp$D2(0) case studied in the previous section, the
D2$\perp$D2(0) supergravity solution has the A-branes initially
smeared over $z_b$.  The field theory on the intersection is still
$0+1$ dimensional, and so we conclude that the B-branes delocalize
over the full near-A-horizon spacetime.  In considering the
implications for the asymptotically flat solutions, we recall that the
near-A-horizon region has been enlarged by smearing over $z_b$.  The
figure below shows the result for three clumps of type A branes placed
close enough to each other that their charge radii ($r_2$, indicated
by dotted lines) overlap.  The full near-horizon region is the
interior of the solid heavy line.  Thus, blowup modes near the center
(thin solid line) are now allowed to be much larger while remaining
inside the near-horizon region.  When enough clumps are present to
model a complete smearing of the type A branes, we may expect the
field theory prediction to imply complete delocalization in the $z_a$
directions of the type B branes in the asymptotically flat solution as
well as the near-A-horizon solution.

\vskip-0.0truein\hskip1.0truein{\epsfbox{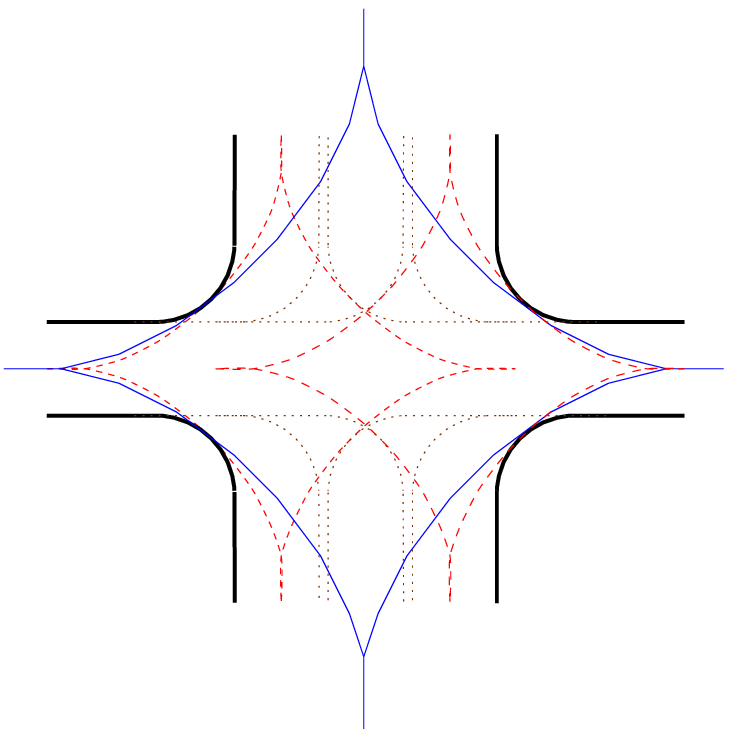}}

We draw entirely analogous conclusions for the other orthogonally
intersecting cases which have $d=0+1$ or $d=1+1$ dimensional
intersections, such as the D1$\perp$D3(0).  The only case which can
localize from the field theory perspective has a $d=2+1$ intersection,
and this is D4$\perp$D4(2).  But since this has only three totally
transverse coordinates it exerts a $1/r$ potential and its
near-horizon supergravity solution is localized, so again we have
agreement of the near-horizon supergravity and field theory.
Following our above argument, there will be asymptotically flat
solutions in which both branes are localized, in addition to the known
solutions where the A branes are initially smeared over the $z_b$
directions.

\section{Discussion}
\label{disc}

We have seen that many supergravity solutions containing two types of
branes (A and B) have the property that one of the branes (B)
delocalizes when the transverse separation between the branes is
removed.  This happens when the world volume directions of the type B
brane are contained in the world volume directions of the type A
brane, or in directions in which the type A brane has been smeared,
and when the dimension of the intersection manifold is sufficiently
small.

In terms of the corresponding brane gauge theories, this phenomenon is
associated with the lack of a sharp transition between the `separated'
and `coincident' branches in the limit where the separation between
the branes is very small.  It is also associated with the fact that
the asymptotic values of massless fields do not label superselection
sectors in 0+1 and 1+1 dimensions; i.e., with the
Coleman-Mermin-Wagner theorem \cite{MW,SC}.
In particular, the supergravity delocalization is related to large
fluctuations of some size moduli fields in the `coincident' branch of
the super Yang-Mills theory.  As mentioned in \cite{SM}, the classical
delocalization is closely related to the black hole no-hair
`theorems', recently reviewed in \cite{Bek}.  It is rather amusing to
connect such a classic feature of black hole physics with quantum
fluctuations in the super Yang-Mills theory.  Up to subtleties
discussed in section III, we find agreement both with respect to which
cases should delocalize and with respect to the rate at which this
delocalization occurs as the transverse separation between the branes
is removed.  It would be very interesting to understand in detail
exactly why this rate agrees so well.

We have also seen this delocalization to be in accord with
expectations that there should in fact exist orthogonal intersecting
brane supergravity solutions with {\it both} branes (A and B)
localized in the directions along the other brane.  In this case, our
delocalization phenomenon may become a finite neck of the supergravity
solution of the sort that is seen \cite{BIons} in the Born-Infeld
description of intersecting branes.  Thus, when the intersection
manifold is 0+1 or 1+1 dimensional, we expect only solutions with
necks of some minimum finite size while, for higher dimensional cases
we expect solutions with necks of all finite sizes (including zero).

There were, however, some cases that we were not able to analyze
properly.  In some of these cases, the type A brane has been smeared
so that it covers a 6+1 dimensional volume.  The corresponding
spacetime then resembles, to a certain extent, that of a D6-brane and
the brane gauge theory may not properly decouple from the bulk.  We
were therefore unable to rely on holography to draw conclusions about
the classical supergravity from the quantum gauge theory.
Nonetheless, we found agreement for the D2$\parallel$D6 and
D4$\perp$D4(2) cases, and we would like to understand why this
happened.

As mentioned in \cite{SM}, BPS supergravity solutions for many three-
(and higher)-charge solutions can also be analyzed in this way.
Typically, when two of the charges are smeared, we can discuss
localization of the third just as was done for the type B branes
above.  Asymptotically flat situations of this type that involve only
Ramond-Ramond branes include three sets of D$2$ branes, or three sets
of D$3$ branes.  In these cases, the branes are again smeared over a
6+1-dimensional volume and we do not expect decoupling from the bulk.

Other cases that could not be studied precisely include NS-NS objects.
Consider first the case of fundamental strings intersecting R-R
branes.  Here, we are stymied by our lack of understanding of
fundamental strings in R-R backgrounds\footnote{For recent progress in
this direction, see \cite{RRB1,RRB2,RRB3}.}.  However, in the
supergravity regime, either the curvature in string units or the
dilaton becomes large near the core of the R-R branes.  This suggests
that the fundamental string will fluctuate significantly near the
supergravity R-R branes, and that this should give rise to
delocalization of the endpoint (necking) near the core of the R-R
branes in analogy with our discussion of solutions describing
D-strings intersecting a D3-brane.

For the case of D$p$-branes intersecting NS5-branes, where the
intersection has $p-1$ space and one time dimension, we have little to
say because we do not understand well enough the theory on the
NS5-branes or the related theory on the intersection manifold.  It may
be described by some sector of the NS5 `little string theory'
\cite{LST1,LST2}, but such a description is likely to require much more
than a field theory.  Note that there are only $3$ directions
transverse to both branes for any value of $p$, so that the
supergravity solutions, with one brane smeared over the other, will
always be localized.

There remains however, the annoying case of D4$\parallel$D0 with the
D4-branes smeared over one $w$ direction.  Here, our classical
supergravity and quantum field theory analyses predict delocalization.
But, despite the fact that the potential is $r^{-2}$ and we expect
holography to hold, the two descriptions disagree with regard to the
rate at which this should happen.  It appears that smearing just the
D4 branes is a more subtle operation in the quantum brane gauge theory
than the supergravity would have us believe.

\acknowledgements
We would like to thank Andr\`es Gomberoff, Akikazu Hashimoto, David
Kastor, Andy Strominger, Jennie Traschen, and especially Clifford
Johnson and Nikita Nekrassov, for useful discussions.
In addition, we are extremely grateful to Joe Polchinski both for his
patience and his valuable insight, and for explanations of several
matters integral to Section IIIB. This work was supported in part by
National Science Foundation grant No. PHY94-07194.  D.M was supported
in part by NSF grant No.  PHY97-22362 and funds from Syracuse
University.

\appendix

\section{A comparison with near-core solutions}
\label{Y}
One potential confusion concerns the near-core solutions of
\cite{Youm}.  While they are valid solutions to the supergravity
equations, we now argue that, due to the subtleties of boundary
conditions, they are not the appropriate ones to consider in our
context.

To discuss those solutions, it is necessary to recall some elements of
their construction.  The differential equations to be solved are just
(\ref{A}) and (\ref{B}) in the regime $\hat r_A \gg r_\perp, r_{\perp
0}$ under the assumption that no $w$ directions (associated with
smearing of the type A branes in directions transverse to the type B
branes) are present.  In \cite{Youm}, $r_{\perp 0}$ was set to zero.
However, we will find it useful to keep $r_{\perp 0}$ nonzero and then
study the limit in which it vanishes.  Following \cite{Youm}, we take
the $z_{a}$ directions to be non-compact.  We take the number of
$z_{a}$ directions to be $D$.  We would like to have spherical
symmetry in both the $x_\perp$ and $z_{a}$ directions so, as in
section \ref{SUGRAV}, we replace the fully localized type B brane with
a spherical ($S^{d-1} \times S^{D-1}$) shell of source: $q_B
r_\perp^{-(d-1)} r_{a}^{-(D-1)} \delta (r_\perp, r_{\perp 0}) \delta
(r_{a}, r_{a 0})$.  Thus, we have spherical symmetry in both the
$z_{a}$ and $x_{\perp}$ directions and the solution depends only on
$r_\perp$ and $r_{a} = |z_{a}|$.

In the region $\hat r_A \gg r_\perp, r_{\perp 0}$, \cite{Youm} uses a
trick first introduced in \cite{ITY} and finds that the equation
simplifies under a change of coordinates (correcting a typographic
error in
\cite{Youm}):
\begin{equation}
\label{trans}
r_{\perp} \rightarrow Y = {{2 \hat r_A^{(d-2)/2}} \over {|4-d|}}
r_\perp^{{4-d} \over 2}.
\end{equation}
In terms of $Y$, the equation to be solved (for $\hat r_A \gg r_\perp,
r_{\perp 0}$) may be written
\begin{eqnarray}
Y^{-d/(4-d)}\left( \partial_Y( Y^{d/(4-d)} \partial_Y
H_B) \right) &+& r_{a}^{1-D}\left(  \partial_{r_{a}} (
r_{a}^{D-1} H_B ) \right ) \cr &=&
Y^{-d/(4-d)} r_{a}^{1-D} \delta(Y - Y(r_{\perp 0}))
\delta(r_{a},r_{a0}).
\end{eqnarray}
This equation may be solved by realizing that it is the analytic
continuation (to non-integer dimensions) of Laplace's equation on
$\Rl^{( {d \over {4-d}} +1) } \times \Rl^{D}$, in coordinates in which
the $SO({d \over 4-d} +1) \times SO(D)$ symmetry is manifest.  The
closed form solutions of \cite{Youm} may be obtained by smearing out
the source further over the $S^{{d \over {d-4}} + D}$ sphere $Y^2 +
r_{a}^2 = R^2_0$ and taking the limit as $R_0 \rightarrow 0$.  What is
important to note is that, for $d > 4$, the coordinate transformation
(\ref{trans}) means that $R_0 \rightarrow 0$ corresponds to taking
$r_{\perp 0}$ to infinity.  Thus, the type B and A branes are not in
fact being placed at the same location in space.  One may verify that,
near the type A branes, the solutions of \cite{Youm} for $d > 4$ do
not depend on the $z_{a}$ coordinates, and so in this sense are not
localized solutions.  The fact that, at a generic point in the
spacetime, the solutions of \cite{Youm} do depend on $z_{a}$ is a
reflection of the lack of an asymptotically flat region in the
near-core spacetime: even though the type B branes have been taken to
infinity, part of their field can still be seen.  For the case $d=4$,
the coordinate transformation (\ref{trans}) breaks down, but
\cite{Youm} constructs a logarithmic solution to which similar remarks
apply.

We note that if one tries to use these methods to construct a
localized solution (for $d>4$) by taking $r_{a0}, r_{\perp 0}
\rightarrow 0$, one takes the source to infinity in $Y$ where it will
have little impact.  Thus, this seems to reproduce the conclusion of
section \ref{SUGRAV} that the type B branes delocalize.  However, the
analysis is complicated by the nontrivial mapping of surfaces in
$z_{a}, x_{\perp}$ space into $z_{a},Y$ space.

\end{document}